\documentclass[11pt, a4paper]{article}
\usepackage{authblk} 
\usepackage{svg}
\usepackage{siunitx} 
\usepackage{float}
\usepackage{graphicx}
\usepackage{dcolumn}
\usepackage{bm}
\usepackage[colorlinks=true, allcolors=blue]{hyperref}

\newcommand{\email}[1]{\thanks{Electronic address: \href{mailto:#1}{#1}}}
\newcommand{\ack}{\section*{Acknowledgments}}
\newcommand{\funding}[1]{\bigskip \noindent \textbf{Funding:} #1}
\newcommand{\roles}[1]{\bigskip \noindent \textbf{Author Contributions:} #1}
\newcommand{\data}[1]{\bigskip \noindent \textbf{Data Availability Statement:} #1}

\title{Wafer-Scale Micro-Knife Sealed Vacuum Cells for Quantum Devices}

\author{Megan Lauree Kelleher$^{1,*}$, Konrad Ziegler$^1$, Jeremy Robin$^1$, Lianxin Huang$^1$, Mitchel Button$^1$, Liam Mauck$^1$, Judith Olson$^1$, Peter Brewer$^1$, Danny Kim$^1$, John Kitching$^2$, Ruwan Senaratne$^1$, William R McGehee$^2$ and Travis M Autry$^{1,*}$}

\affil{$^1$Materials and Microsystems Laboratory, HRL Laboratories, 3011 Malibu Canyon Road, Malibu, CA 90265-4797, USA}
\affil{$^2$Time and Frequency Division, National Institute of Standards and Technology, 325 Broadway, Boulder, CO 80305, USA}
\affil{$^*$Authors to whom any correspondence should be addressed.}

\begin{document}
\maketitle

\email{mlkelleher@hrl.com, tmautry@hrl.com}

\begin{abstract}
Advanced integration technologies greatly enhance the prospects and reliability of practical quantum sensors, atomic clocks, and quantum information technologies. The performance and proliferation of these devices at chip scale is contingent upon developing low leak and low gas permeation vacuum cells using wafer-scale techniques.  Here, we demonstrate a novel, low-leak-rate micro-knife bonding approach, enabling the realization of both atomic vapor cells and more complex evacuated atomic beam devices. These devices are fabricated using selective laser etching in a fused silica platform. The vapor cells are mechanically robust exhibiting shear-force strength $\sim 15$MPa, demonstrate long lifetimes ($> 1$ year), low residual gas pressures $ (\ll 10^{-3} \, \text{mbar}) $, and leak rates below fine-leak testing sensitivity ($\ll 2.8 \times 10^{-10} \frac{\text{mbar} \cdot \text{L}}{\text{s}}$). Micro-knife bonding greatly simplifies the fabrication process for complex chip scale atom-beam devices and atomic vapor cells while identifying a path to future chip-scale cold atom devices, improved chip scale atomic clocks, and fieldable dissipation-dilution-limited optomechanics.
\end{abstract}


\section{\label{sec:level1}Introduction}

	The promise of chip-scale quantum devices for  quantum information, precision sensing, and timekeeping is contingent upon the ability to realize compact portable devices that do not degrade due to their environment. Significant progress has been realized for warm atomic clocks and magnetometers with the adoption of wafer-scale fabrication of rough-vacuum and pressurized cells for confinement of atomic vapors \cite{kitchingChipscaleAtomicDevices2018}. Future advanced atomic sensors containing either trapped ion \cite{schulzSidebandCoolingCoherent2008,kaufmannThickfilmTechnologyUltra2012,jauLowpowerMiniature171Yb2012,schwindtHighlyMiniaturizedVacuum2016,sivernsIonTrapArchitectures2017,hankinSystematicUncertaintyDue2019,raggSegmentedIontrapFabrication2019a,auchterIndustriallyMicrofabricatedIon2022a,reensHighFidelityIonState2022}  or neutral atom \cite{gallegoOpticalLatticeAtom2009,rushtonFeasibilityFullyMiniaturized2014,straatsmaOnchipOpticalLattice2015,mcgilliganGratingChipsQuantum2017,mcgeheeMagnetoopticalTrappingUsing2021,isichenkoPhotonicIntegratedBeam2023b,hekiHighEfficiencyLargeangle2024} chips require ultra-high-vacuum (UHV) to avoid background gas collisions and enable laser cooling. The supporting integrated atom-photonic elements  \cite{mehtaIntegratedOpticalAddressing2016,mcgilliganMicrofabricatedComponentsCold2022,lohOperationOpticalAtomic2020a,chauhanVisibleLightPhotonic2021b,lohOpticalAtomicClock2025,bregazziSingleBeamGratingChip3D2025a}, especially clock lasers, exhibit temperature sensitivities that will necessitate vacuum sealing for environment control \cite{lohUltranarrowLinewidthBrillouin2019,mclemoreMiniaturizingUltrastableElectromagnetic2022b,joWaferlevelHermeticallySealed2022,quackIntegratedSiliconPhotonic2023}.  Finally, advanced dissipation dilution limited (Q $\sim 10^{9}$) opto-mechanical devices and superconducting qubits also benefit from ultra-high-vacuum sealing for improved coherence \cite{capellePolarimetricAnalysisStress2017,tsaturyanUltracoherentNanomechanicalResonators2017,ghadimiElasticStrainEngineering2018,mergenthalerUltrahighVacuumPackaging2021}. 

Conventional wafer-scale techniques for hermetic sealing are sufficient for capacitively readout micro-electro-mechanical systems (MEMS) that require vacuum levels of $\sim 1$ mbar (100 Pa, Q \(\sim 10^{4}\)) vacuum and are tolerant of sealing temperatures of $400$\textdegree C - $950$\textdegree C \cite{torunbalciGoldTinEutectic2014,torunbalciMethodWaferLevel2015,asadianUltrahighVacuumPackaging2017b,chenRobustMethodFabricating2017,torunbalciAllSiliconProcessPlatform2021,joWaferlevelHermeticallySealed2022}.  In contrast, the effective leak rate leads to degraded performance for quantum sensors \cite{straessleLowtemperatureIndiumbondedAlkali2013a,dellisLowHeliumPermeation2016a,kitchingChipscaleAtomicDevices2018,martinCompactOpticalAtomic2018,lemkeMeasurementOpticalRubidium2022,carleReductionHeliumPermeation2023a} and prevents the realization of ultra-high-vacuum devices on chip. The effective leak rate consists of two contributions, the bond-interface leak rate and the gas-permeation rate through material substrates. These leaks are often due to the prevalence of glass wafers for optical transparency or SiO\textsubscript{2} thin films used for optical access, integrated photonic claddings, and electrical routing. The prevailing method for sealing atomic vapor cells is anodic bonding, a method with leak rates comparable to hermetic military standards $\sim$ ($\SI{1e-8}{\frac{mbar\cdot L}{s}}, \SI{1e-6}{\frac{Pa\cdot L}{s}} $) \cite{akinRFTelemetryPowering1990,liewMicrofabricatedAlkaliAtom2004,zhangMicrofabricationHermeticityMeasurement2019}.  However, future UHV chip-scale quantum devices require leak rates many orders of magnitude lower ($ < \SI{1e-19}{\frac{mbar\cdot L}{s}} $) \cite{rushtonFeasibilityFullyMiniaturized2014}.  Thus both current and future sensors would benefit from improved sealing methods exhibiting lower interfacial leak rates while also enabling the use of unconventional He-impermeable single-crystalline and transparent materials such as sapphire and silicon carbide \cite{rushtonFeasibilityFullyMiniaturized2014,mcgilliganMicrofabricatedComponentsCold2022}.  
		
 \begin{figure*}
\includegraphics[width=\textwidth]{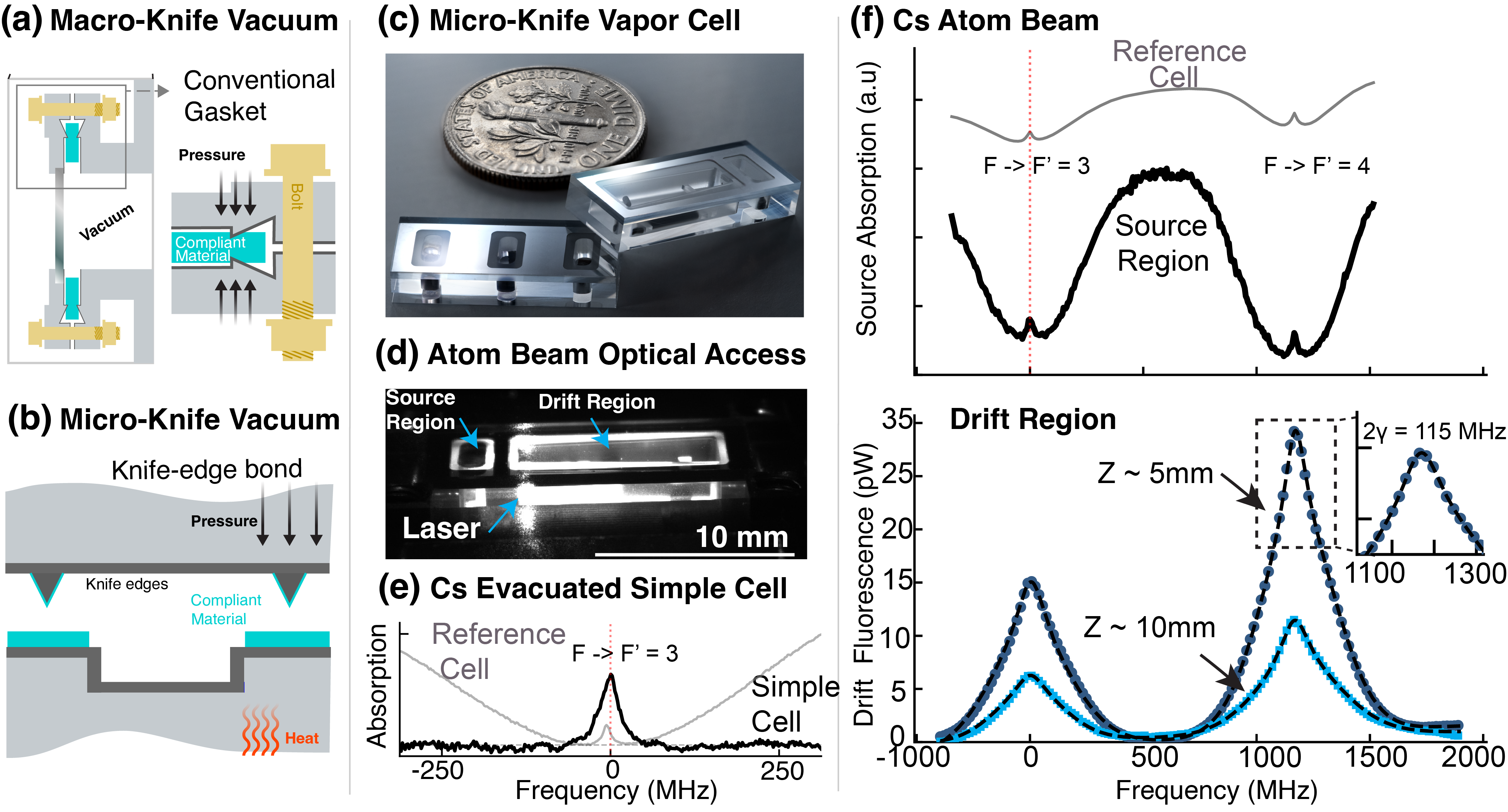}
\caption{\label{fig:deform} (a) A conventional extreme-high-vacuum chamber cuts into a compliant metal gasket and pressure is maintained with a mechanical bolt. (b) A micro-knife evacuated vacuum chamber is realized by plastically deforming and diffusion bonding two metals together. (c) Close up photograph of a fabricated all fused silica atom-beam and simple atomic vapor cells with a dime for scale.  (d) All-glass vacuum cells provide a high-degree of optical access enabling background free fluorescence measurements. (e) Background free saturated absorption spectrum of the D$1$ line ($895$ nm) from an evacuated Cs vapor cell.  Some residual gas causes slight broadening. (f) (Top) Spectra taken of a Cs atom beam in the source region immediately after activation. (Bottom) Spectra taken from the atom beam in the drift region using side excitation shows Cs collimation.  The dashed line represents a fitted spectrum and shows a beam spectral FWHM of $\sim 115$ MHz. For panels (d-f) cell temperature is between $80$ \textdegree C - $90$ \textdegree C.  Saturated absorption and fluorescence spectra are measured using different experimental setups.}
\end{figure*}		

	We draw inspiration from experiments involving laser cooled neutral atoms and ions operating in actively pumped, ultra-high-vacuum chambers sealed with knife-edge seals. These knife edge seals enable extreme-high-vacuum (XHV) and the incorporation of different materials in lab-scale experiments. Conventional knife-edges fabricated on steel vacuum chambers plastically deform—or cut—soft metal gaskets (Fig. \ref{fig:deform}a). Held in place with bolts, these knife edge interfaces realize intimate contact between the steel knife and the gasket achieving measurement-limited ultra-low leak rates ($ \ll \SI{1e-11}{\frac{mbar\cdot L}{s}} $) \cite{elsenLeakageRateDetection2021}. 
	
	Plastic deformation \textit{bonding} has been demonstrated as a viable technique for lowering the temperature of encapsulation in MEMS vacuum systems, exhibits high yield, and exhibits some of the lowest leak rates in the literature at $ < \SI{1e-14}{\frac{mbar \cdot L}{s}} $ for single-seal rings \cite{anteliusSmallFootprintWaferlevel2011,wangWaferLevelVacuumPackaging2017,wangWaferLevelVacuumSealing2019}.  At the micro-scale, plastic deformation bonding has been restricted to silicon due to its micro-machinability and excess shearing forces in glass \cite{anteliusSmallFootprintWaferlevel2011}. In our new approach, we demonstrate a direct analog to macro-scale vacuum seals by realizing micro-knife deformation bonding and demonstrate its potential for diverse substrates  (Fig.  \ref{fig:deform}b).
	
	Our chip-scale atomic vapor cells are fabricated using an unconventional substrate,  fused silica. Selective laser etching of fused silica wafers is used to form cavities \cite{luciveroLaserwrittenVaporCells2022,artusio-glimpseWaferlevelFabricationAlldielectric2025} and micro-capillaries for evacuated atom-beam cells and vapor cells \cite{liRobustCharacterizationMicrofabricated2020,weiCollimatedVersatileAtomic2022,martinezChipscaleAtomicBeam2023}. Metal deposition is used to realize high pressure contact points/knives. These titanium knives are bonded to a thick compliant metal layer. Two prototypical all-glass chip designs are demonstrated (Fig. \ref{fig:deform}c). The first design consists of a chip containing three evacuated simple vapor cells and the second design consists of an evacuated atom-beam cell.  An advantage of this technique is the optical transparency of the side walls, which enables interrogation of the vapor cell along multiple directions; side excitation (Fig. 1d)
and vertical collection provides background free fluorescence measurements. Conventional microfabricated vapor cells use etched cavities in Si capped with glass and are restricted to a single optical axis \cite{kitchingChipscaleAtomicDevices2018}.  Cell leak rates were measured using $^{85}$Kr fine leak testing and are measurement-limited to $\ll 2.8 \times 10^{-10} \frac{\text{mbar} \cdot \text{L}}{\text{s}}$. 
		
This new process involves only a single bonding interface, dramatically reducing fabrication complexity, as conventional simple vapor cells require two bonds, while complex atom-beam devices currently require four bonds. With a reduction in complexity comes a corresponding improvement in yield that, based on our current results, transitions our atom-beam prototypes \cite{martinezChipscaleAtomicBeam2023}—previously bonded at the individual die level—to a wafer-scale process. While yield is not the focus of this work,  a $> 85\,\%$ yield of Cs devices at wafer scale was achieved,  limited primarily by the Cs pill source.

 \begin{figure*}
\includegraphics[width=\textwidth]{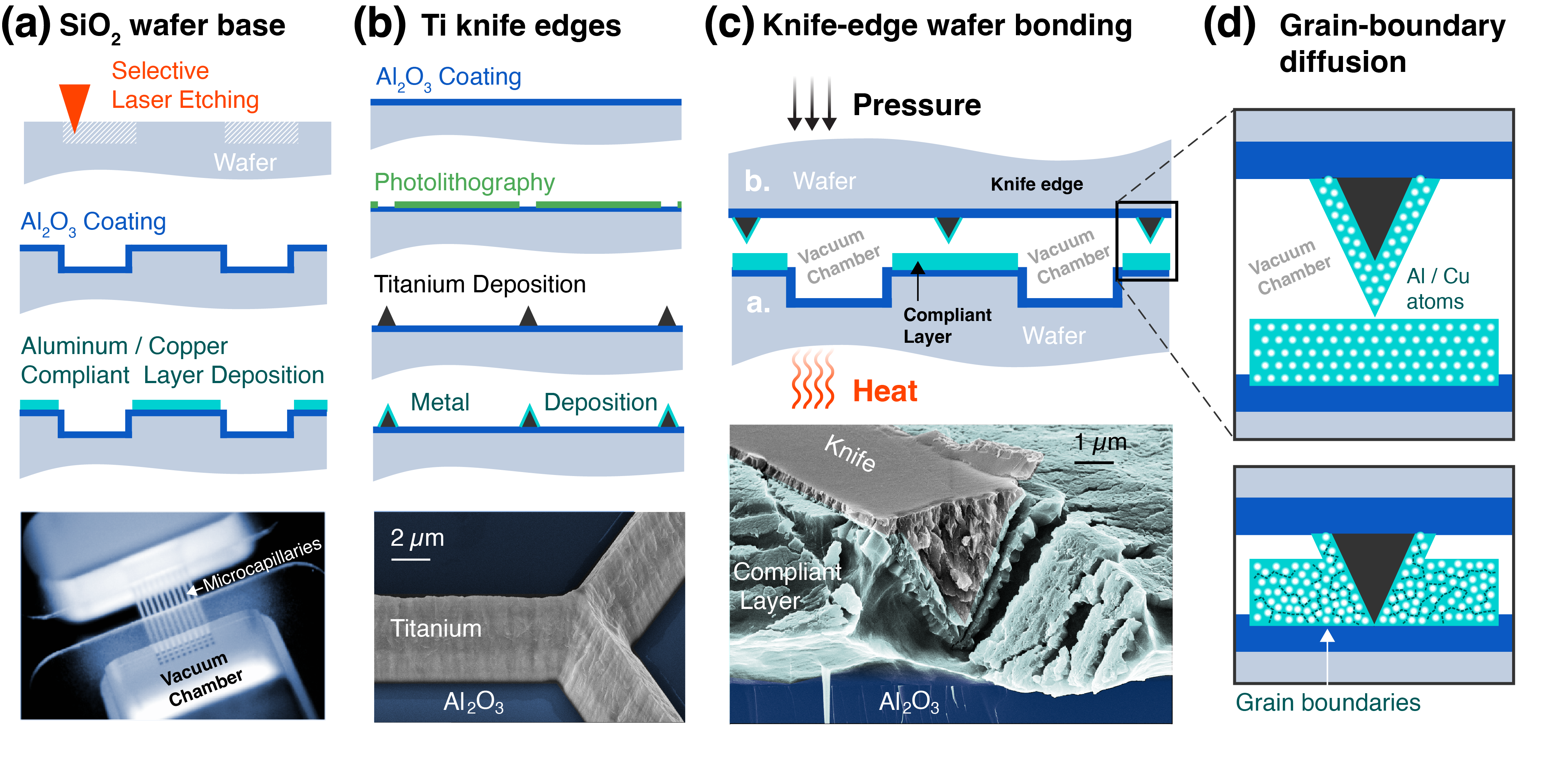}
\caption{\label{fig:Vapor_Cell_Fab} (a) Internal cavities and other structures (such as micro-capillary arrays) are fabricated using selective laser etching.  A coating consisting of Al$_2$O$_3$ is deposited to reduce He permeation.  A thick compliant metal layer is then deposited to form sealing surfaces. Inset showing the vertically stacked micro-capillaries $ (50 \, \mu\text{m} \times 75 \, \mu\text{m} \times 1.8 \, \text{mm}) $ patterned in a single step with the cavity to connect the source region and the drift region in an atom beam cell. (b) A capping wafer is fabricated by first depositing Al$_2$O$_3$ and then depositing knives followed by a capping/bonding layer of metal. Inset showing a knife junction in a honeycomb seal. (c) After outgassing, the wafers are brought together and bonded. Inset: Scanning electron microscope (SEM) image showing a cleaved interface. The knife is bonded into a compliant layer with some damage from cleaving such that the capping wafer has detached from the knife bottom. (d) Diffusion at the grain boundary between the two layers creates a hermetic seal and provides mechanical strength.}
\end{figure*}

\section{\label{sec:level2} Spectroscopy of Vapor Cells}

		Saturated absorption spectra of evacuated simple cells and atom beam cells on the D$1$ line ($895\,$nm) were used to characterize the vacuum cells and Cs background gas.  Simple cells consistently show narrow sub-Doppler peaks with widths $\sim 10\text{--}20$~MHz. Some residual gas is present in simple cells providing slight broadening in background free spectroscopy $\sim 10$ MHz (Fig. \ref{fig:deform}e) attributed to insufficient getter capacity, incomplete getter activation, and source outgassing when activated. 
		
		 In contrast to the simple vapor cells, our evacuated atom beams cells exhibit very low residual gas pressure characteristic of a well-collimated atomic beam \cite{ramseynormanMolecularBeams1956}.  Typical spectra taken post-activation in the source region of an atom beam cells also show sub-Doppler peaks (Fig. \ref{fig:deform}f - top). Additionally, fluorescence spectra taken in the drift region show atom beam collimation.  The atom beam spectral shape is empirically fitted to a Gaussian background and a Lorentzian (Fig. \ref{fig:deform}f - bottom).  This measurement shows a collimated atom beam width (HWHM) of $\gamma \sim 57$ MHz in the most central region of the fluorescence peak. To form an atomic beam,  the background gas pressure must be sufficiently low to support molecular flow and avoid diffusive flow. Therefore,  the residual gas pressure can be inferred to be $\ll 10^{-3} \, \text{mbar}$ since the mean free path of the Cs vapor supports atomic collimation at $\sim 10$ mm.  This calculation assumes $\text{Cs-He or Cs-N}_2$ collisions dominate providing a conservative estimate of background pressure based on collisional cross-sections \cite{estermannMeanFreePaths1947a}. 
				
\section{\label{sec:level3} Vacuum Cell Fabrication With Plastic Deformation Bonding}

	Ideal bonding conditions include pristine, ultra-flat, angstrom level roughness, and that ultra-clean wafers be brought into contact at high temperatures. For flat-flat bonding of wafers, voids occur due to imperfections in wafer flatness and the presence of particulates \cite{rebhanPhysicalMechanismsCoppercopper2015b,hinterreiterSurfacePretreatedLowtemperature2018}. This problem is exacerbated with heavily processed wafers, such as those with deep cavities for MEMS devices and atomic vapors, where surface quality is compromised. Beyond surface quality issues, high temperature processing is needed to overcome bonding barriers for fusion bonding and metal-metal bonding alike. For instance, conventional MEMS bonding metals (aluminum, copper) readily form oxides that act as diffusion barriers, thus requiring high temperatures and surface preparation to achieve robust bonds \cite{rebhanPhysicalMechanismsCoppercopper2015b,hinterreiterSurfacePretreatedLowtemperature2018,karlenSealingMEMSAtomic2020}.

	To sidestep these processing issues, we employ metal-metal plastic deformation bonding with micro knife edges in a wafer-scale process (Fig. \ref{fig:Vapor_Cell_Fab}). Plastic deformation occurs when the applied compressive pressure on a material exceeds the material's ability to respond linearly via Hooke's law and permanent deformation of the structure occurs. The deformation process is continuous but materials exhibit critical points known as the yield-point where significant deformation begins. For deposited metals, these points are process dependent, however they are typically  $\sim 200\,$MPa \cite{wangWaferLevelVacuumPackaging2017}.  Deformation is realized by the application of localized pressure with our micro-fabricated knives realizing ultra-high pressure (GPa) contact points. 

	Fabrication begins with preparing a cavity wafer through selective laser etching a fused silica wafer. This process enables the 3D etching of fused silica through the writing of acid-selective strain/nanogratings and has been used to demonstrate atomic vapor cells  \cite{luciveroLaserwrittenVaporCells2022}. This paper extends this process to the case of more complex atom-beam devices \cite{liCascadedCollimatorAtomic2019,liRobustCharacterizationMicrofabricated2020,weiCollimatedVersatileAtomic2022,martinezChipscaleAtomicBeam2023} with in-situ fabrication of a 3D array of horizontal microcapillaries with dimensions 50 $\mu$m x $75 \mu$m x $1.8$ mm (Fig. \ref{fig:Vapor_Cell_Fab}a). Laser etching of internal structures enables the realization of atomic vapor cells and atom-beam collimator arrays with only a single bonding interface dramatically improving the process complexity compared to conventional devices. The processed cavity wafer is given a coating of $\mathrm{Al}_2\mathrm{O}_3$ before a thick $\sim 1\,\mu\text{m}$--$10\,\mu\text{m}$ layer of metal is evaporated to form a compliant-deformation layer. The cavity wafer is loaded with both Cs pills (CS/AMAX/PILL/1-$0.6$) and non-evaporable getter pills (ST-$172$) from SAES. Atom beam cells are additionally loaded with graphite rods for differential pumping of alkali vapor \cite{martinezChipscaleAtomicBeam2023}. 

	A separate capping wafer (Fig. \ref{fig:Vapor_Cell_Fab}b) is given a coating of $\mathrm{Al}_2\mathrm{O}_3$ before $\sim 1 \,\mu\text{m}$--$10\,\mu\text{m}$ tall Ti knives are fabricated. These knives naturally converge to very sharp $\sim 10\,\text{nm}$ - $50\,\text{nm}$ sized tips and deposited Ti has $\sim 10$ GPa hardness \cite{arshiThicknessEffectProperties2013}.  These knives are subsequently patterned with an additional metal overlay layer (Cu/Al) for diffusion bonding.

\begin{figure}[t] 
\centering
\includegraphics[width=.7\columnwidth]{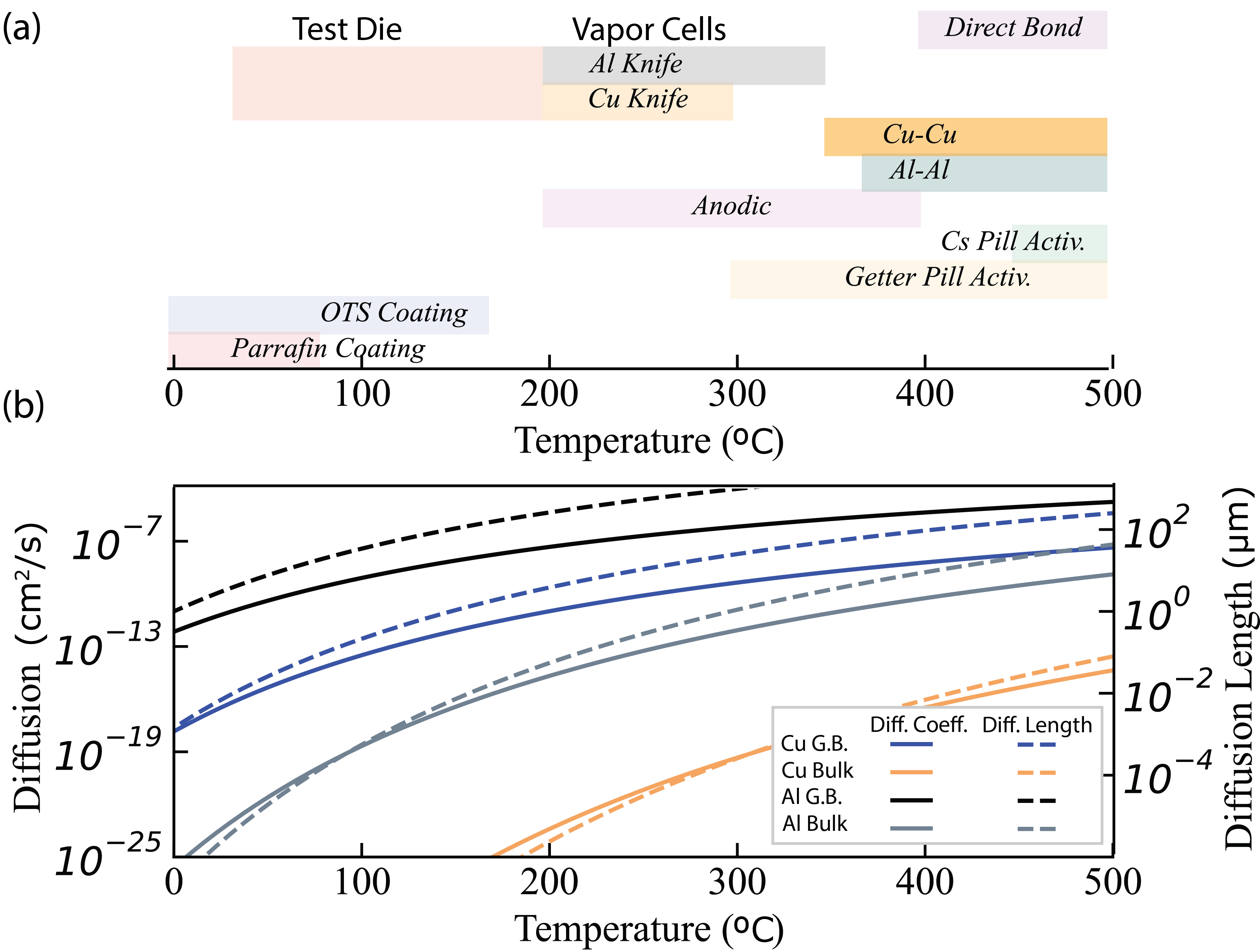}
\caption{\label{fig:process_windows} (a) Process interdependency for atomic vapor cell devices including realized bonds, pill activation, and relevant coatings. Shown are typical direct (fusion) bonding conditions, anodic bonding conditions, and metal-metal thermo-compression bonding conditions for flat wafers. Test bonds (not vapor cells) at 40~\textdegree C show that ultra-low temperature bonding is possible. (b) Calculated diffusion constant and diffusion length for Cu and Al for bulk and grain boundaries (G.B.). The exact value for this plot may vary depending on parameters such as grain boundary size.}
\end{figure}

	The cap and cavity wafers are outgassed and then brought into contact, causing deformation and bonding to fresh metal (Fig. \ref{fig:Vapor_Cell_Fab}d). The bond employed here is that of a grain-boundary thermo-compression (diffusion) bond.   The inclusion of plastic deformation lowers the necessary bonding temperature compared to conventional metal-metal thermo-compression bonds.  Effectively, intimate contact (Fig. \ref{fig:process_windows}a) between the knife and compliant layer bypasses the stringent requirements on surface oxides, surface residue, and wafer flatness as it forms metal-metal bonds with fresh sub-surface metal.
 
  The diffusion coefficient and diffusion length at grain boundaries are much larger than within a bulk material, lowering the requisite bonding temperature (Fig. \ref{fig:process_windows}b).
 
	Low-temperature bonding is particularly crucial for chip-scale atomic devices with a complex interdependency (Fig. \ref{fig:process_windows}a). Although hermetically sealed devices are best outgassed at high temperature, getters such as SAES ST172 activate at $\sim 300$\textdegree C and most atomic sources would exhibit significant mass loss from a high temperature bakeout. This point is critical, as with the exception of temperature-tolerant pills and azides for Rb and Cs, other chip-scale atomic sources contain only nanograms of material and are loaded in their pure form \cite{manginellSituDissolutionDeposition2012,pickLowpowerMicrostructuredAtomic2025,kumarFastresponseLowPower2025}. Moreover, cell-wall coatings like paraffin are incompatible with high-temperature bonding or bakeouts.  Test bonds at $40$\textdegree C with the micro-knife bonding process yielded, but vapor cells were not fabricated at this temperature. 
 			
\begin{figure}
    \centering 
    \includegraphics[width=.5\columnwidth]{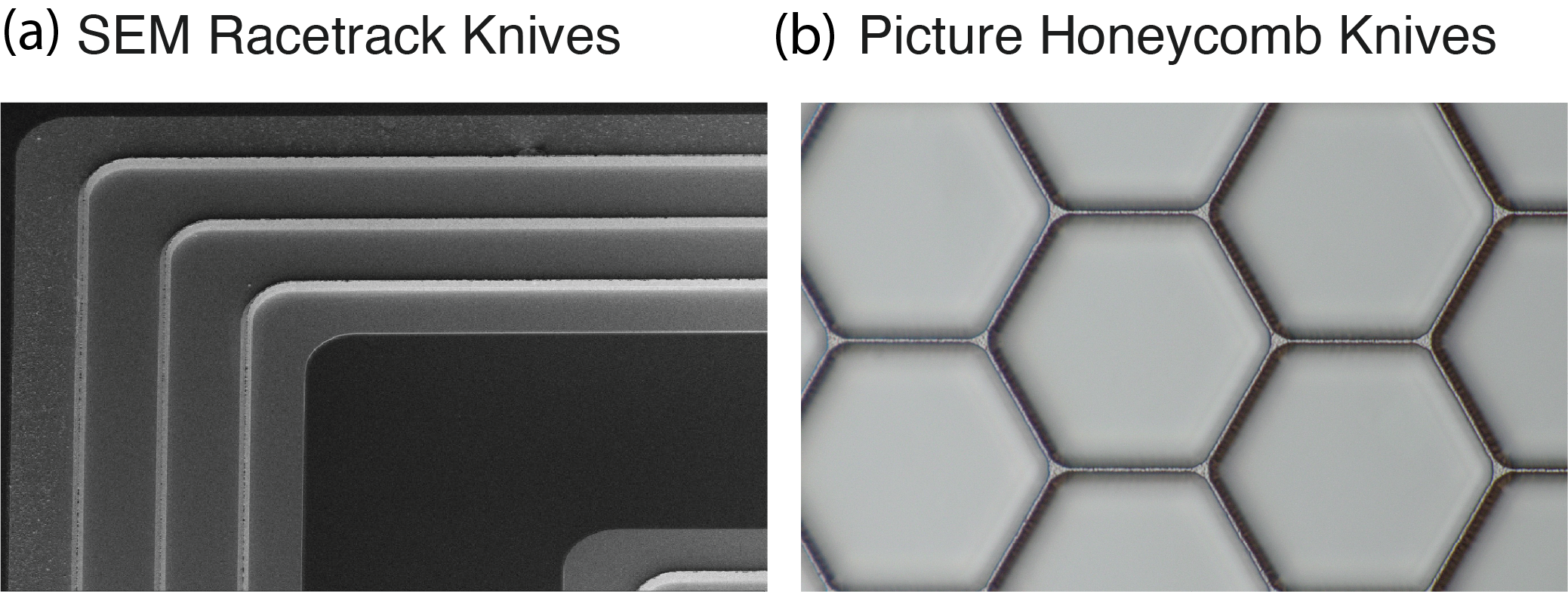}
    \caption{\label{fig:seals} 
    (a) SEM image of a racetrack knife showing vacuum moats. (b) Picture of patterned honeycomb seals
showing unconventional vacuum moats. Note: Images are from different devices and test bonds.}
\end{figure}
			 
\section{\label{sec:level5}Seal Design, Mechanical Strength, Yield}

	To achieve ultra-low-leak-rate-capable devices, the knife edge mask incorporates a few features.  First,  nested racetrack knife-edges (Fig. \ref{fig:seals}a) provide conventional vacuum moats (nested seals) to reduce overall system leak rate  \cite{rushtonFeasibilityFullyMiniaturized2014}. Second, the devices include an unconventional vacuum moat in the form of a honeycomb lattice knife (Fig. \ref{fig:seals}b). The devices also include an $\mathrm{Al_2O_3}$ coating on both the cell cap and vapor cell chamber which is reported to reduce He permeation 
\cite{woetzelLifetimeImprovementMicrofabricated2013a,dellisLowHeliumPermeation2016a}. An atom-beam cell was sent out for Kr-$85$ fine leak testing and was found to be below the sensitivity limit of $2.8 \times 10^{-10} \frac{\text{mbar} \cdot \text{L}}{\text{s}}$.  Future ultra-fine leak testing with impermeable all-silicon cavities may determine the single-seal and moat-seal leak rate \cite{anteliusSmallFootprintWaferlevel2011,wangWaferLevelVacuumPackaging2017,wangWaferLevelVacuumSealing2019}.

	 The vapor cell yield is primarily limited by the activation step of the Zr-Al-Cs pills from SAES. The pills flake and break during manual cell loading. According to discussions with SAES, pill flaking indicates pill poisoning due to water; a likely result of the high relative humidity in Malibu, CA. Despite these issues, a yield of $>85\,\%$ of the working cells across the wafer was realized.
	 
	The activation of the NEG and Cs pills is done with a high power 980 nm laser and a $\sim 1$ mm spot size. For successfully activated devices window darkening can typically be observed when the activation laser exceeds $\sim 2$ W of power on the Cs pills. Cells made with and without the Al$_2$O$_3$ coating are observed to have a qualitative difference in cell darkening (improved with the coating) but darkening remained apparent. 
		 
	During activation of each vapor cell the presence of Cs atomic vapor is monitored with a resonant $895$ nm laser and a reference saturated absorption spectroscopy vapor cell. After activation, devices are taken to a different setup where light is sent into the vapor cells through their transparent sidewalls.  In this setup, background free fluorescence is acquired. 
		 
 \begin{figure}
 \centering
\includegraphics[width=.5\columnwidth]{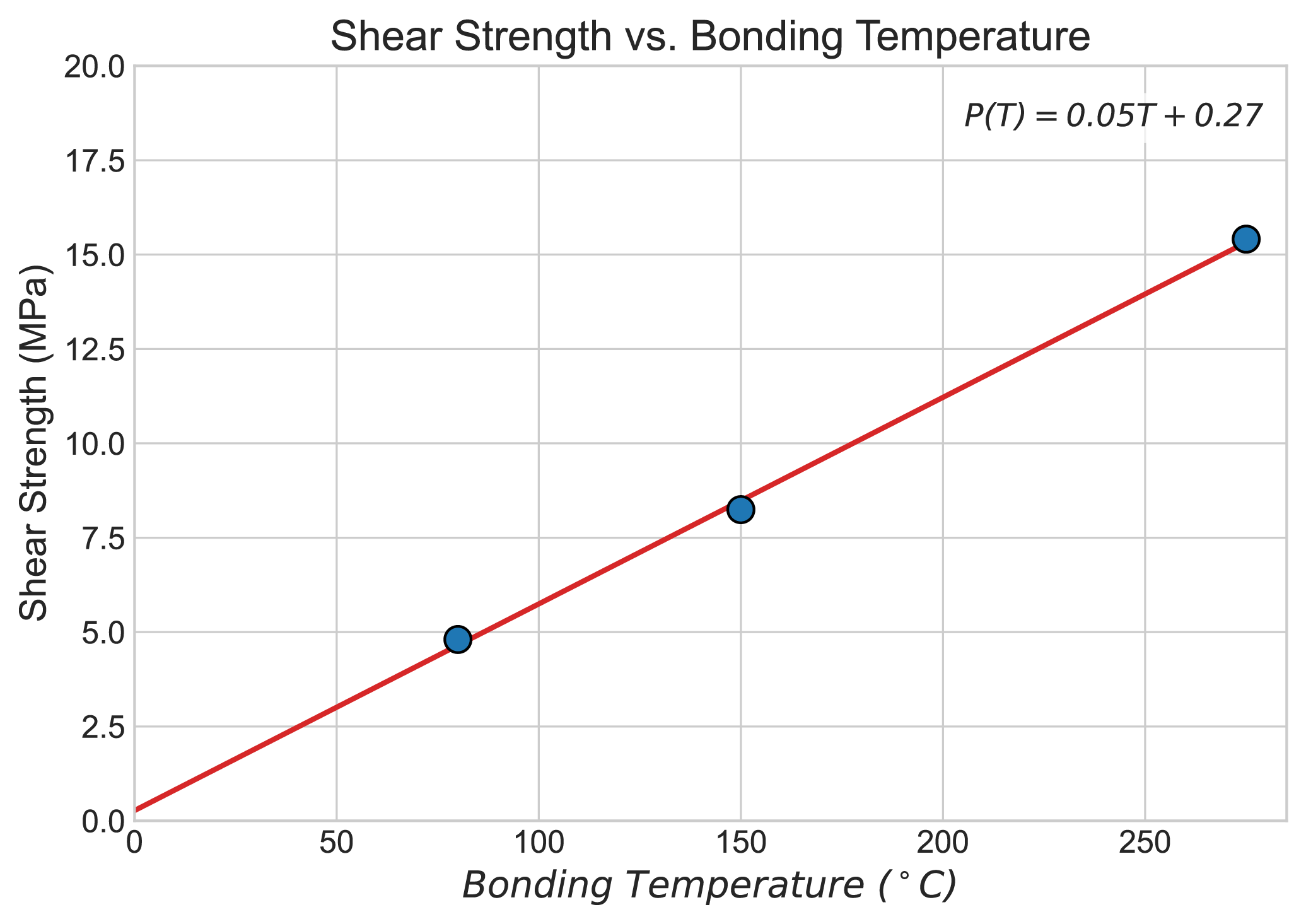}%
\caption{\label{fig:shear_force} Shear force testing shows the knife-edge bond strength is linear with increasing temperature.  A maximum bond strength of $\sim 15$MPa is observed.}
\end{figure}

	 For the simple evacuated vapor cells, devices exhibit no decrease in atomic vapor pressure over a year. However, window darkening can create further complication for atom-beam cells as partial-activation of Cs can lead to device lifetimes on the orders of hours instead of weeks as expected for the amount of Cs available in a pill $\sim .5$ mg. Typically, atom-beam cells are activated with monitoring of an absorption signal in the source region. If the absorption signal decreases after ten minutes, the cell is deemed partially activated and re-activated at a slightly higher power.  Most often full activation of the beam cells results in window darkening.
	 	 
	 	 Shear force testing of the bond strength was performed on fabricated test die. The die was mounted using epoxy to metal blocks and sheared along the bonding interface. For different bonding temperatures there is a linear dependence on the bonding strength with temperature (Fig. \ref{fig:shear_force}). For the lowest temperature bonded test die ($40$ \textdegree C) the epoxy used to mount the device caused it to debond and fail. This confirms that ultra-low temperature bonding, while possible, will require extensive work to realize yielding hermetic seals. 
		 		 	 
\section{\label{sec:level6} Conclusion}
	We have demonstrated a new wafer-scale bonding process for realizing evacuated vapor cells. This process demonstrates several new features. First, we demonstrate that a micro knife edge plastic deformation process can realize long lasting hermetic vapor cells. Second, we show the utility and fabrication simplicity enabled by selective laser etching and use this process to realize complex vapor cell geometries like atom-beam cells with a single bonding interface at wafer-scale.  This new process already provides ultra-low leak rates and is expected to support the bonding of single-crystalline transparent materials and the wafer-scale realization of ultra-high vacuum cells for laser cooling of atoms and dissipation-dilution limited optomechanical devices. Beyond atomic vapor cells, our process supports hermetically sealing photonic integrated circuits for improved stability and the $3$D integration of vapor cells above photonic integrated circuits.	

\ack{The authors acknowledge John Carpenter and Luke Quezada for assistance in making figures. Tracy Boden, John Carlson, and Bruce Holden are acknowledged for their help in fabricating devices. Any mention of commercial products is for information only; it does not imply recommendation or endorsement by NIST or HRL. Any opinions, findings and conclusions or recommendations expressed in this material are those of the author(s) and do not necessarily reflect the views of the Defense Advanced Research Projects Agency (DARPA).}

\funding{This material is based upon work supported by the Defense Advanced Research Projects Agency (DARPA) under Contract No. HR0011-23-C-0042 as part of the H6 program. Distribution Statement "A" (Approved for Public Release, Distribution Unlimited).}

\roles{M.L.K. and T.M.A. conceived the experiments. M.L.K., K.Z., J.R., L.H., M.B., L.M., P.B., D.K., and R.S. contributed to device fabrication and characterization. J.O. provided technical advice. J.K., and W.R.M. provided technical guidance on experiments and device design. T.M.A. supervised the program and designed the devices. All authors reviewed the manuscript.}

\data{The data that support the findings of this study are available from the corresponding author upon reasonable request.}	 
\bibliographystyle{unsrt} 
\bibliography{bonding_updatedv7}
	
\end{document}